\documentclass[preprint,12pt]{elsarticle}

\pdfoutput=1

\usepackage{amssymb,amsmath}
\usepackage{bm}
\usepackage{graphicx}

\biboptions{sort&compress}

\begin{document}

\begin{frontmatter}


\title{MNPBEM -- A Matlab toolbox for the simulation of plasmonic nanoparticles}

\author{Ulrich Hohenester}
\ead{ulrich.hohenester@uni-graz.at}
\ead[url]{http://physik.uni-graz.at/~uxh}
\author{Andreas Tr\"ugler}
\address{Institut f\"ur Physik. Karl--Franzens--Universit\"at Graz, 
  Universit\"atsplatz 5, 8010 Graz, Austria}



\begin{abstract}
\texttt{MNPBEM} is a Matlab toolbox for the simulation of metallic nanoparticles (MNP), using a boundary element method (BEM) approach.  The main purpose of the toolbox is to solve Maxwell's equations for a dielectric environment where bodies with homogeneous and isotropic dielectric functions are separated by abrupt interfaces. Although the approach is in principle suited for arbitrary body sizes and photon energies, it is tested (and probably works best) for metallic nanoparticles with sizes ranging from a few to a few hundreds of nanometers, and for frequencies in the optical and near-infrared regime.  The toolbox has been implemented with Matlab classes. These classes can be easily combined, which has the advantage that one can adapt the simulation programs flexibly for various applications.

\end{abstract}


\end{frontmatter}

\section*{Program summary}

\noindent{\sl Program title:} \texttt{MNPBEM}.\\
\noindent{\sl Programming language:} Matlab 7.11.0 (R2010b)\\
\noindent{\sl Computer:} Any which supports Matlab 7.11.0 (R2010b)\\
\noindent{\sl Operating system:} Any which supports Matlab 7.11.0 (R2010b)\\
\noindent{\sl RAM required to execute with typical data:} $\ge 1$ GByte\\
\noindent{\sl Has the code been vectorised or parallelized?:} no\\
\noindent{\sl Keywords:} Plasmonics, metallic nanoparticles, boundary element method\\
\noindent{\sl CPC Library Classification:} Optics\\
\noindent{\sl External routines/libraries used:} \texttt{MESH2D} available at \texttt{www.mathworks.com}\\
\noindent{\sl Nature of problem:} Solve Maxwell's equations for dielectric particles with homogeneous dielectric functions separated by abrupt interfaces\\
\noindent{\sl Solution method:}Boundary element method using electromagnetic potentials\\
\noindent{\sl Running time:} Depending on surface discretization between seconds and hours\\

\section{Introduction}\label{sec:intro}

Plasmonics is an emerging field with numerous applications foreseen, ranging from sensorics over extreme light concentration and light harvesting to optical and quantum technology, as well as metamaterials and optical cloaking \cite{atwater:07,maier:07,pendry:06,schuller:10,novotny:11}.  The workhorse of plasmonics are \textit{surface plasmons}, these are coherent electron charge oscillations bound to the interface between a metal and a dielectric \cite{maier:07,novotny:06}.  These surface plasmons come along with strongly localized, so-called evanescent electromagnetic fields, which can be exploited for bringing light down to the nanoscale, thereby overcoming the diffraction limit of light and bridging between the micrometer length scale of optics and the nanometer length scale of nanostructures.  On the other hand, tiny variations of the dielectric environment close to the nanostructures, e.g. induced by binding of molecules to a functionalized metal surface, can significantly modify the evanescent fields and, in turn, the surface plasmon resonances.  This can be exploited for (bio)sensor applications, eventually bringing the sensitivity down to the single-molecule level.  

Of particular interest are \textit{particle plasmons}, these are surface plasmons confined in all three spatial dimensions to the surface of a nanoparticle \cite{kreibig:95,maier:07}.  The properties of these excitations depend strongly on particle geometry and interparticle coupling, and give rise to a variety of effects, such as frequency-dependent absorption and scattering or near field enhancement.  Particle plasmons enable the concentration of light fields to nanoscale volumes and play a key role in surface enhanced spectroscopy \cite{chance:78,kneipp:97,nie:97,anger:06}.  

Simulation of particle plasmons is nothing but the solution of Maxwell's equations for metallic nanoparticles embedded in a dielectric environment.  Consequently, the simulation toolboxes usually employed in the field are not specifically designed for plasmonics applications.  For instance, the discrete dipole approximation toolbox \texttt{DDSCATT} \cite{draine:88,draine:94} was originally designed for the simulation of scattering from interstellar graphite grains, but has in recent years been widely used within the field of plasmonics.  Also the finite difference time domain (FDTD) approach \cite{yee:66,ward:00} has been developed as a general simulation toolkit for the solution of Maxwell's equations.  Other computational approaches widely used in the field of plasmonics are the dyadic Green function technique \cite{martin:95} or the multiple multipole method \cite{novotny:06}.

In this paper we present the simulation toolbox \texttt{MNPBEM} for metallic nanoparticles (MNP), which is based on a boundary element method (BEM) approach developed by Garcia de Abajo and Howie \cite{garcia:02,garcia:10}.  The approach is less general than the above approaches, in that it assumes a dielectric environment where bodies with homogeneous and isotropic dielectric functions are separated by abrupt interfaces, rather than allowing for a general inhomogeneous dielectric environment.  On the other hand, for most plasmonics applications with metallic nanoparticles embedded in a dielectric background the BEM approach appears to be a natural choice.  It has the advantage that only the boundaries between the different dielectric materials have to be discretized, and not the whole volume, which results in faster simulations with more moderate memory requirements.  

The \texttt{MNPBEM} toolbox has been designed such that it provides a flexible toolkit for the simulation of the electromagnetic properties of plasmonic nanoparticles.  The toolbox works in principle for arbitrary dielectric bodies with homogeneous dielectric properties, which are separated by abrupt interfaces, although we have primarily used and tested it for metallic nanoparticles with diameters ranging from a few to a few hundred nanometers, and for frequencies in the optical and near-infrared regime.  We have developed the programs over the last few years \cite{hohenester.prb:05,hohenester.ieee:08}, and have used them for the simulation of optical properties of plamonic particles \cite{truegler.prb:08,truegler.prb:11}, surface enhanced spectroscopy \cite{gerber.prb:07,reil:08,koller:10}, sensorics \cite{becker:10,jakab:11}, and electron energy loss spectroscopy (EELS) \cite{schaffer.prb:09,hohenester.prl:09}.  

In the past year, we have completely rewritten the code using classes within Matlab 7.11.  These classes can be easily combined such that one can adapt the simulation programs flexibly to the user's needs.  A comprehensive help is available for all classes and functions of the toolbox through the \texttt{doc} command.  In addition, we have created detailed help pages, accessible in the Matlab help browser, together with a complete list of the classes and functions of the toolbox, and a number of demo programs.  In this paper we provide an ample overview of the \texttt{MNPBEM} toolbox, but leave the details to the help pages.  As the theory underlying our BEM approach has been presented in great detail elsewhere \cite{garcia:02,garcia:10}, in the following we only give a short account of the approach and refer the interested reader to the pertinent literature and the help pages.

Throughout the \texttt{MNPBEM} toolbox, lengths are measured in nanometers and photon energies through the light wavelength $\lambda$ (in vacuum) in nanometers.  In the programs we use for $\lambda$ the notation \verb|enei| (inverse of photon energy).  With the only exception of the classes for the dielectric functions, one could also measure distances and wavelengths in other units such as e.g. micrometers or atomic units.  Inside the toolbox we use Gauss units, in accordance with Refs.~\cite{garcia:02,garcia:10}.  This is advantageous for the scalar and vector potentials, which are at the heart of our BEM approach, and which could not be treated on an equal footing with the SI system.  For most applications, however, the units remain completely hidden within the core routines of the BEM solvers.

\section{Theory}\label{sec:theory}

\begin{figure}
\centerline{%
  \includegraphics[width=0.3\columnwidth]{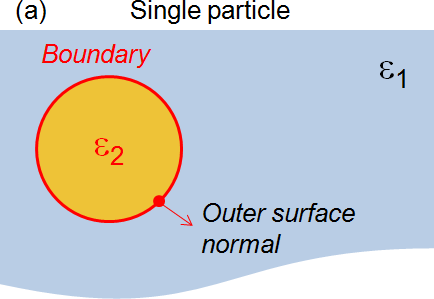}\quad
  \includegraphics[width=0.3\columnwidth]{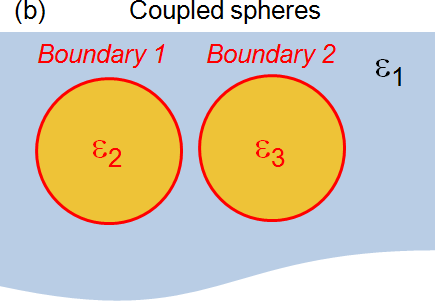}\quad
  \includegraphics[width=0.3\columnwidth]{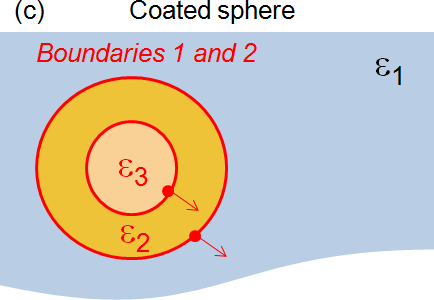}}
\caption{A few representative model systems suited for simulation within the \texttt{MNPBEM} toolbox:  (a) Metallic nanosphere embedded in a dielectric background, (b) coupled nanospheres, and (c) coated nanosphere.  The dielectric functions are denoted with $\varepsilon_i$.  In panels (a) and (c) we also report the outer surface normals of the particle boundaries.}
\end{figure}

In the following we consider dielectric nanoparticles, described through local and isotropic dielectric functions $\varepsilon_i(\omega)$, which are separated by sharp boundaries $\partial V_i$.  A few representative examples are shown in Fig.~1.  When the particles are excited by some external perturbation, such as an incoming plane wave or the fields created by a nearby oscillating dipole, they will become polarized and electromagnetic fields are induced.  The goal of the \texttt{MNPBEM} toolbox is to compute for a given external perturbation these induced electromagnetic fields.  This is achieved by solving Maxwell's equations and using the boundary conditions at the particle boundaries.

\subsection{Quasistatic approximation}

We first discuss things for nanoparticles much smaller than the light wavelength, where one can employ the quasistatic approximation.  Here one solves the Poisson or Laplace equation for the electrostatic potential \cite{jackson:99}, rather than the Helmholtz equation for the scalar and vector potentials, but keeps the full frequency-dependent dielectric functions in the evaluation of the boundary conditions \cite{novotny:06,hohenester.ieee:08}.  A convenient solution scheme is provided by the electrostatic Green function
\begin{equation}\label{eq:greenstat}
  \nabla^2 G(\bm r,\bm r')=-4\pi\delta(\bm r-\bm r')\,,\quad
  G(\bm r,\bm r')=\frac 1{|\bm r-\bm r'|}\,,
\end{equation}
which is the proper solution of the Poisson equation for a point-like source in an unbounded, homogeneous medium.  In case of an inhomogeneous dielectric environment, with homogeneous dielectric particles $V_i$ separated by sharp boundaries $\partial V_i$, inside a given region $\bm r\in V_i$ one can write down the solution in the \textit{ad-hoc} form \cite{garcia:02,garcia:10,hohenester.prb:05}
\begin{equation}\label{eq:adhoc}
  \phi(\bm r)=\phi_{\rm ext}(\bm r)+\oint_{V_i} G(\bm r,\bm s)
  \sigma(\bm s)\,da\,.
\end{equation}
Here $\phi_{\rm ext}$ is the external electrostatic potential, and $\sigma(\bm s)$ is a surface charge distribution located at the particle boundary $\partial V_i$.  Eq.~\eqref{eq:adhoc} is constructed such that it fulfills the Poisson or Laplace equation everywhere except at the particle boundaries.  The surface charge distribution $\sigma(\bm s)$ has to be chosen such that the appropriate boundary conditions of Maxwell's equations are fulfilled.  Continuity of the parallel electric field implies that $\sigma$ has to be the same in- and outside the particle.  From the continuity of the normal dielectric displacement we find the boundary integral equation \cite{fuchs:75,hohenester.prb:05,garcia:10}
\begin{equation}\label{eq:bim}
  \Lambda\,\sigma(\bm s)+\oint \frac{\partial G(\bm s,\bm s')}%
  {\partial  n}\sigma(\bm s)\,da'=-
  \frac{\partial\phi_{\rm ext}(\bm s)}{\partial n}\,,\quad
  \Lambda=2\pi\frac{\varepsilon_2+\varepsilon_1}%
  {\varepsilon_2-\varepsilon_1}\,,
\end{equation}
whose solutions determine the surface charge distribution $\sigma$.  Here $\frac\partial{\partial n}$ denotes the derivative along the direction of the outer surface normal, and $\varepsilon_1$ and $\varepsilon_2$ are the dielectric functions in- and outside the particle boundary, respectively.  Approximating the integral in Eq.~\eqref{eq:bim} by a sum over surface elements, 
one arrives at a \textit{boundary element method} (BEM) approach with surface charges $\sigma_i$ given at the discretized surface elements.  From
\begin{equation}\label{eq:bem}
  \Lambda\,\sigma_i+\sum_j \left(\frac{\partial G} {\partial  n}\right)_{ij}
  \sigma_j=-\left(\frac{\partial\phi_{\rm ext}}{\partial n}\right)_i
\end{equation}
one can obtain the surface charges $\sigma_i$ through simple matrix inversion.  Equation~\eqref{eq:bem} constitutes the main equation of the quasistatic BEM approach.  The central elements are $\Lambda$, which is governed by the dielectric functions in- and outside the particle boundaries, the surface derivatives $\left(\frac{\partial G}{\partial n}\right)_{ij}$ of the Green function connecting surface elements $i$ and $j$, and the surface derivative $\left(\frac{\partial\phi_{\rm ext}}{\partial n}\right)_i$ of the external potential.

\subsection{Full Maxwell equations}

A similar scheme can be applied when solving the full Maxwell equations.  We now need both the scalar and vector potentials, which both fulfill a Helmholtz equation \cite{jackson:99,garcia:02}.  Again we define a Green function
\begin{equation}\label{eq:greenret}
  \left(\nabla^2+k_i^2\right)G_i(\bm r,\bm r')=-4\pi\delta(\bm r-\bm r')\,,\quad
  G_i(\bm r,\bm r')=\frac{e^{ik_i|\bm r-\bm r'|}}{|\bm r-\bm r'|}\,,
\end{equation}
where $k_i=\sqrt{\varepsilon_i}k$ is the wavenumber in the medium $\bm r\in V_i$, $k=\omega/c$ is the wavenumber in vacuum, and $c$ is the speed of light.  The magnetic permeability $\mu$ is set to one throughout.  In analogy to the quasistatic case, for an inhomogeneous dielectric environment we write down the solutions in the \textit{ad-hoc} form \cite{garcia:02,garcia:10}
\begin{eqnarray}
  \phi(\bm r)&=&\phi_{\rm ext}(\bm r)+
    \oint_{V_i} G_i(\bm r,\bm s)\sigma_i(\bm s)\,da\label{eq:adhocphi}\\
  \bm A(\bm r)&=&\bm A_{\rm ext}(\bm r)+
    \oint_{V_i} G_i(\bm r,\bm s)\bm h_i(\bm s)\,da\,, \label{eq:adhoca}
\end{eqnarray}
which fulfill the Helmholtz equations everywhere except at the particle boundaries.  $\sigma_i$ and $\bm h_i$ are surface charge and current distributions, and $\phi_{\rm ext}$ and $\bm A_{\rm ext}$ are the scalar and vector potentials characterizing the external perturbation. 

The scalar and vector potentials are additionally related through the Lorentz gauge condition $\nabla\cdot\bm A=ik\varepsilon\phi$ \cite{garcia:02}, which in principle allows to express $\phi$ through the divergence of $\bm A$.  However, it is advantageous to keep both the scalar and vector potential: in the evaluation of the boundary conditions we then only need the potentials together with their first, rather than also second, surface derivatives.  The potential-based BEM approach then invokes only $G_i$ together with its surface derivative $\partial G_i/\partial n$, in contrast to the field-based BEM approach which also invokes the second surface derivatives $\partial^2 G_i/\partial n^2$ \cite{jackson:99,chew:95}.  As higher derivatives of $G_i$ translate to functions with higher spatial variations, it is computationally favorable to keep only first-order surface derivatives.  Another advantage of Eq.~\eqref{eq:adhoca} is that the different components of $\bm A$ are manipulated \textit{separately}.  Thus, when transforming to a BEM approach, by discretizing the boundary integrals, we end up with matrices of the order $N\times N$, where $N$ is the number of boundary elements.  In contrast, for field based BEM approaches the matrices are of the order $3N\times 3N$, where the factor of three accounts for the three spatial dimensions.

In the BEM approach the boundary integrals derived from Eqs.~(\ref{eq:adhocphi},\ref{eq:adhoca}) are approximated by sums over boundary elements.  Exploiting the boundary conditions of Maxwell's equations, in analogy to the quasistatic case, we derive a set of rather lengthy equations for the surface charges and currents \cite{garcia:02,garcia:10} (see also the help pages of the \texttt{MNPBEM} toolbox), which can be solved through matrix inversions and multiplications.  In contrast to the quasistatic case, the surface charges $\sigma_{1,2}$ and currents $\bm h_{1,2}$ in- and outside the boundary (measured with respect to the surface normal $\bm n$) are not identical.  Once $\sigma$ and $\bm h$ are determined, we can compute through Eqs.~(\ref{eq:adhocphi},\ref{eq:adhoca}) the potentials everywhere else, as well as the electromagnetic fields, which are related to the potentials through the usual relations $\bm E=ik\bm A-\nabla\phi$ and $\bm H=\nabla\times\bm A$.

\section{Getting started}\label{sec:start}

\subsection{Installation of the toolbox}

\begin{figure}
\centerline{\includegraphics[width=0.9\columnwidth]{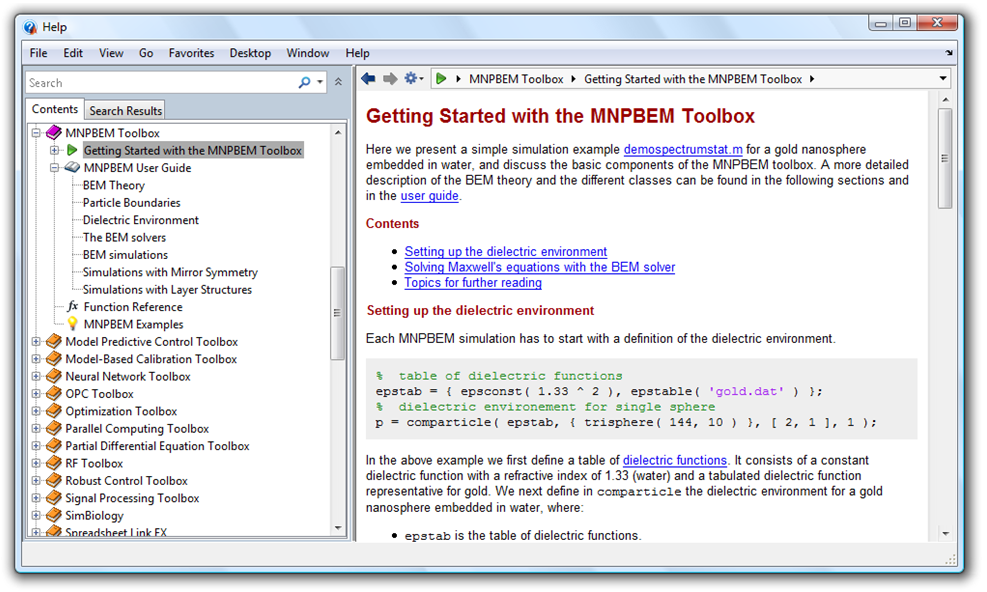}}
\caption{Screenshot of the help pages of the \texttt{MNPBEM} toolbox within the Matlab help browser.  The help pages provide a short introduction, a detailed user guide, a list of the classes and functions of the toolbox, as well as a number of demo programs.}\label{fig:helppages}
\end{figure}

To install the toolbox, one must simply add the path of the main directory \texttt{mnpbemdir} of the \texttt{MNPBEM} toolbox as well as the paths of all subdirectories to the Matlab search path.  This can be done, for instance, through
\begin{verbatim}
addpath(genpath(mnpbemdir));
\end{verbatim}
For particle shapes derived from 2D polygons, to be described in Sec.~\ref{sec:particle}, one additionally needs the toolbox \texttt{MESH2D - Automatic Mesh Generation} available at \texttt{www.mathworks.com}.  Again, one should add the path of the corresponding directory to the Matlab path.

To set up the help pages, one must once change to the main directory of the \texttt{MNPBEM} toolbox and run the program \texttt{makemnpbemhelp}

\begin{verbatim}
>> cd mnpbemdir;
>> makemnpbemhelp;
\end{verbatim}
Once this is done, the help pages, which provide detailed information about the toolbox, are available in the Matlab help browser.  Figure~\ref{fig:helppages} shows a screenshot of the \texttt{MNPBEM} help pages.

\subsection{A simple example}

Let us start with the discussion of a simple example.  We consider a metallic nanosphere embedded in water, which is excited by an electromagnetic plane wave, corresponding to light excitation from a source situated far away from the object.  For this setup we compute the light scattered by the nanosphere.  The file \texttt{demospectrumstat.m} for the corresponding simulation is available in the \texttt{Demo} subdirectory and can be opened by typing 
\begin{verbatim}
>> edit demospectrumstat
\end{verbatim}
The simulation consists of the following steps:

\begin{itemize}
\item[-] define the dielectric functions; 
\vspace*{-0.3cm}\item[-] define the particle boundaries; 
\vspace*{-0.3cm}\item[-] specify how the particle is embedded in the dielectric environment; 
\vspace*{-0.3cm}\item[-] set up a solver for the BEM equations;
\vspace*{-0.3cm}\item[-] specify the excitation scheme (here plane wave excitation); 
\vspace*{-0.3cm}\item[-] solve the BEM equations for the given excitation by computing the auxiliary surface charges (and currents); 
\vspace*{-0.3cm}\item[-] compute the response of the plasmonic nanoparticle (here scattering cross section) to the external excitation.
\end{itemize}

We next discuss the various steps in more detail.  In \texttt{demospectrumstat.m} we first set up a table of dielectric functions, needed for the problem under study, and a discretized particle boundary which is stored in the form of vertices and faces.  A more detailed description of the different elements of the \texttt{MNPBEM} toolbox will be given in the sections below as well as in the help pages.
\begin{verbatim}
%  table of dielectric functions
epstab = {epsconst(1.33^2),epstable('gold.dat')};
%  nanosphere with 144 vertices and 10 nanometers diameter
p = trisphere(144,10);
\end{verbatim}
In the above lines we first define two dielectric functions, one for water (refractive index $n_b=1.33$) and one for gold, and then create a discretized sphere surface with 144 vertices.  For all functions and classes of the toolbox additional information can be obtained by typing
\begin{verbatim}
>> doc trisphere
\end{verbatim}

\begin{figure}
\centerline{\includegraphics[width=0.48\columnwidth]{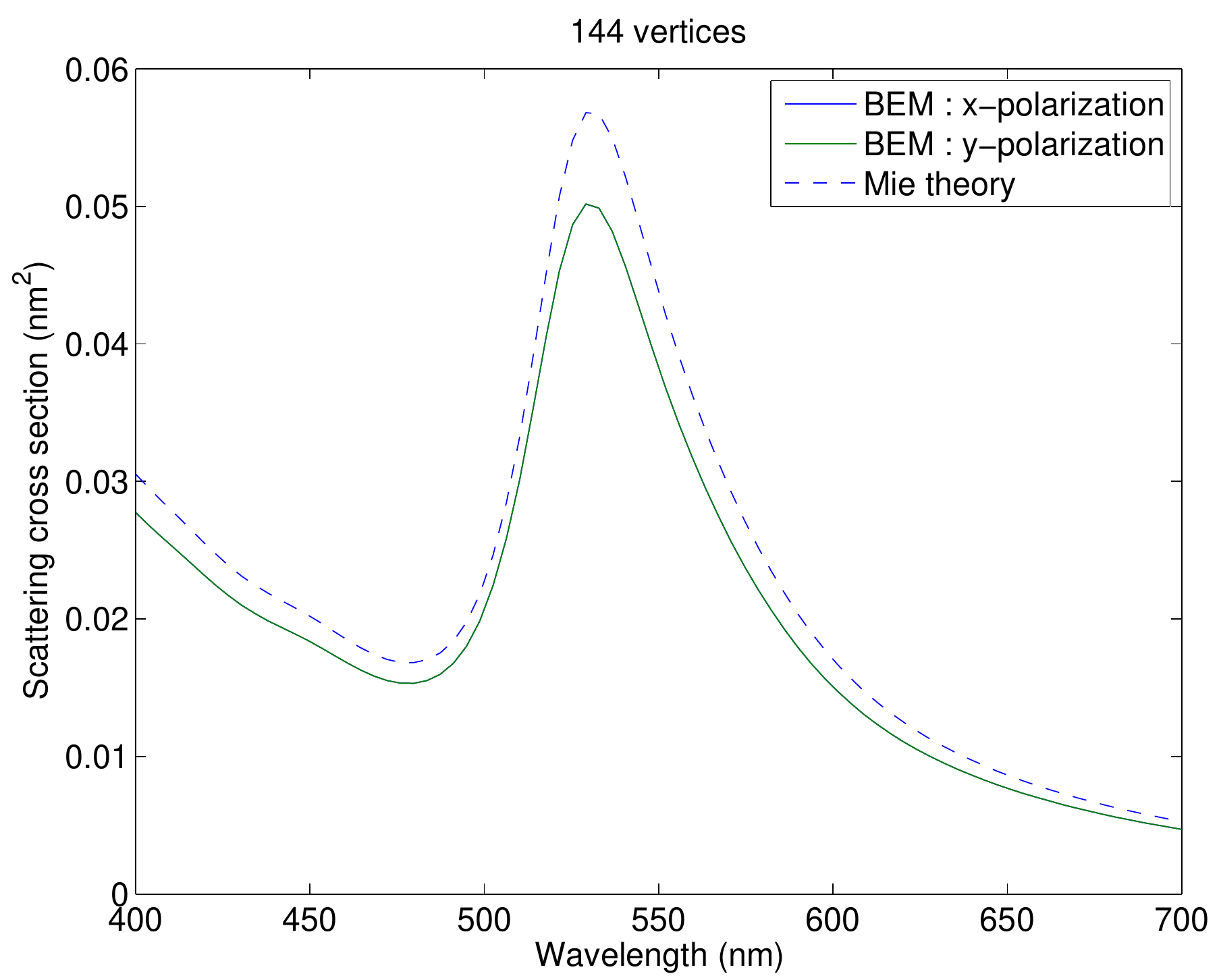}\quad
            \includegraphics[width=0.48\columnwidth]{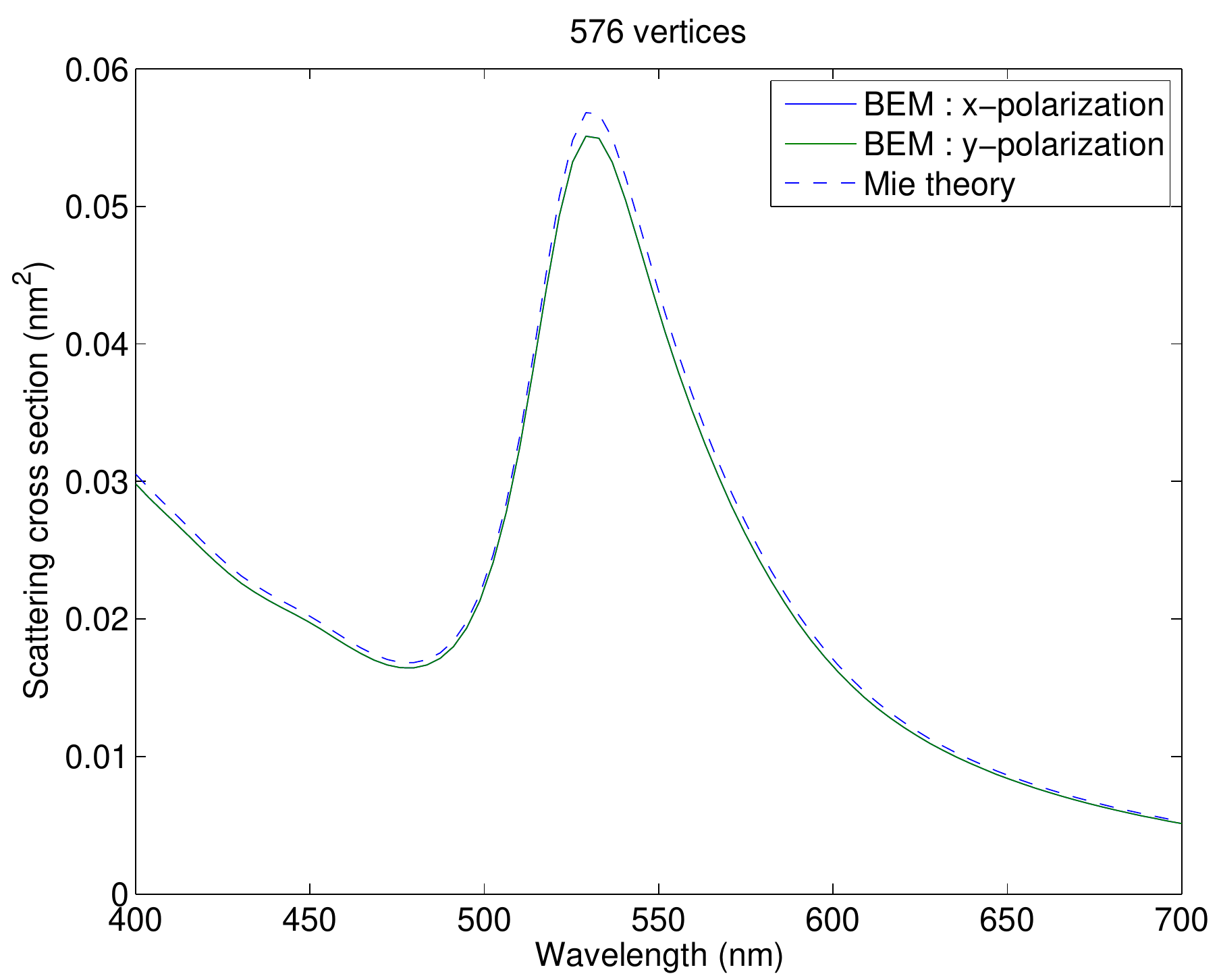}}
\caption{Scattering cross section for a nanosphere with a diameter of 10 nanometers.  We compare the results of BEM simulations for a sphere discretization with 144 (left panel) and 576 (right panel) vertices with those of Mie theory.  The dielectric function of gold is taken from Ref.~\cite{johnson:72} and the refractive index of the embedding medium $n_b=1.33$ is representative for water.  The results for $x$ and $y$ polarization are indistinguishable.}\label{fig:scattering}
\end{figure}

We next have to define the dielectric properties of the nanosphere, depicted in Fig.~1(a), by setting up a \texttt{comparticle} object
\begin{verbatim}
p = comparticle(epstab,{p},[2,1],1);
\end{verbatim}
The first two arguments are cell arrays for the dielectric functions and for the particle boundaries.  The third argument \texttt{inout=[2,1]} describes how the particle boundaries and the dielectric functions are related.  In the above example we specify that the material at the in- and out-side of the boundary are \verb|epstab{2}| and \verb|epstab{1}|, respectively.  Note that the in- and out-side are defined with respect to the surface normal $\bm n$, whose direction is given by the order of the face elements.  To check that these surface normals point into the right direction one can plot the particle with
\begin{verbatim}
>> plot(p,'nvec',true);
\end{verbatim}
Finally, the last argument in the call to \verb|comparticle| indicates that the particle surface is closed.  It is important to provide this additional information, as will be discussed in Sec.~\ref{sec:environment}.  Once the \verb|comparticle| object is set up, it is ready for use with the BEM solvers.  With
\begin{verbatim}
%  quasistatic BEM solver
bem = bemstat(p);
%  plane wave excitation for given light polarizations
exc = planewavestat([1,0,0;0,1,0]);
\end{verbatim}
we set up a solver for the BEM equations within the quasistatic approximation, and a plane wave excitation for polarizations along $x$ and $y$.  We next make a loop over the different wavelengths \verb|enei|.  For each wavelength we solve the BEM equations and compute the scattering cross section \cite{vanhulst:81}
\begin{verbatim}
%  light wavelength in vacuum in nanometers
enei = linspace(400,700,80);
%  scattering spectrum (initialization of array with zeros)
sca = zeros(length(enei),2);

%  main loop over different excitation wavelengths
for ien = 1:length(enei)
  sig = bem \ exc(p,enei(ien));
  sca(ien,:) = exc.sca(sig);
end
\end{verbatim}
The solution of the BEM equations is through \verb|sig=bem\exc(p,enei)|, where \verb|exc(p,enei)| returns for the external light illumination the surface derivative of the potential at the particle boundary.  Finally, we can plot the scattering cross section and compare with the results of Mie theory \cite{vanhulst:81}, Fig.~\ref{fig:scattering}, using the \verb|miestat| class provided by the toolbox.  Similarly, the extinction cross section can be computed with \verb|exc.ext(sig)|.

\begin{figure}
\centerline{\includegraphics[width=0.7\columnwidth]{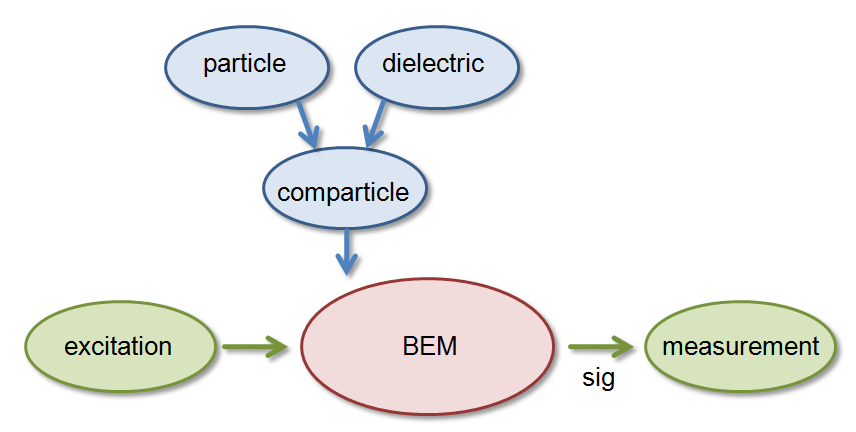}}
\caption{Flow chart for a typical BEM simulations:  we first initialize the BEM solver with a \texttt{comparticle} object, that holds tables of the dielectric functions and of the discretized particle boundaries.  The BEM solver (quasistatic or retarded) communicates with additional excitation-measurement classes, such as \texttt{planewave} for plane wave illumination or \texttt{dipole} for the excitation from an oscillating dipole.  The BEM solver computes the surface charges (currents) \texttt{sig} which can be further processed for measurement purposes.}\label{fig:flow}
\end{figure}

Figure~\ref{fig:flow} shows the flow chart for a typical BEM simulation.  First, we initialize the BEM solver with a \texttt{comparticle} object, that holds tables of the dielectric functions and of the discretized particle boundaries.  The BEM solver then computes for a given excitation the surface charges \verb|sig|, which can then be used for the calculation of measurement results, such as scattering or extinction cross sections.  As the excitation and measurement commands are not hidden inside a function, but appear explicitly inside the wavelength loop, it is possible to further process the results of the BEM simulation.  For instance, we could plot the surface charges through \verb|plot(p,real(sig.sig))| or compute the induced electric fields, as will be described further below.

To compute the scattering cross section for the retarded, i.e. full, Maxwell's equations, we simply have to use a different BEM solver and excitation-measurement class
\begin{verbatim}
%  full BEM solver
bem = bemret(p);
%  plane wave excitation for given light polarizations
exc = planewaveret([1,0,0;0,1,0],[0,0,1;0,0,1]);
\end{verbatim}
Note that in the call to \verb|planewaveret| we now have to specify also the light propagation directions.

\section{Particle boundaries}\label{sec:particle}

The first, and usually most time-consuming job in setting up a \texttt{MNPBEM} simulation is to discretize the particle boundaries.  The discretized boundary is stored as a \verb|particle| object
\begin{verbatim}
p = particle(verts,faces);
\end{verbatim}
Here \verb|verts| are the vertices and \verb|faces| the faces of the boundary elements, similarly to the \verb|patch| objects of Matlab.  \verb|faces| is a $N\times 4$ array with $N$ being the number of boundary elements.  For each element the four entries point to the corners of a quadrilateral.  For triangular boundary elements the last entry should be a \verb|NaN|.

\begin{figure}
\centerline{\includegraphics[width=0.8\columnwidth]{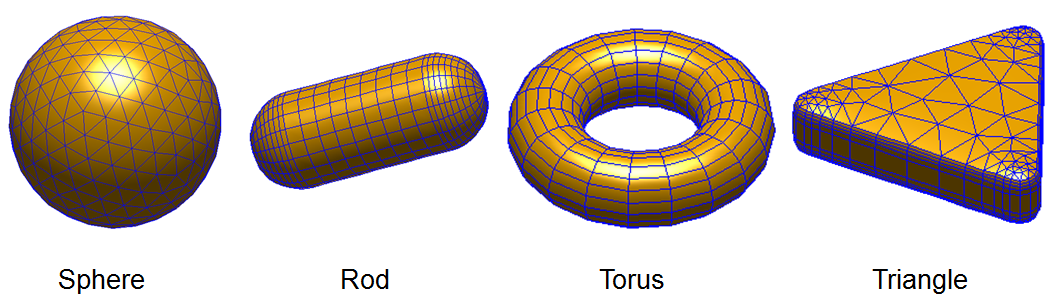}}
\caption{Discretized particle boundaries, as created with the functions \texttt{trisphere}, \texttt{trirod}, and \texttt{tritorus} of the \texttt{MNPBEM} toolbox.  The triangle on the right-hand side is created from a \texttt{polygon} object, which is extruded with the \texttt{tripolygon} function using the \texttt{Mesh2d} toolbox.}\label{fig:particleshapes}
\end{figure}

In principle, in Matlab there exist myriads of surface discretization functions, and the \texttt{MNPBEM} toolbox only provides a few additional ones

\begin{verbatim}
%  sphere with given number of vertices and diameter
p = trisphere(nverts,diameter);
%  nanorod with given diameter and height
p = trirod(diameter,height);
%  torus with given outer and inner radius
p = tritorus(rout,rin);
\end{verbatim}
Figure~\ref{fig:particleshapes} shows the corresponding particle surfaces.  All functions can receive additional options to control the number of discretization points, as detailed in the \verb|doc| command or the help pages.  In addition to the vertices and faces, the \verb|particle| object also stores for each boundary element the area and centroid, as well as three orthogonal vectors, where \verb|nvec| is the outer surface normal.

In many cases one has to deal with flat particles, where a 2D polygon is extruded along the third direction.  The \texttt{MNPBEM} toolbox provides a \verb|polygon| class together with surface discretization and extrusion functions, building upon the \texttt{Mesh2d} toolbox.  Let us look to an example for discretizing the surface of a triangular particle 
\begin{verbatim}
%  2D polygon for triangle with rounded edges
poly = round(polygon(3,'size',[20,20]));
%  height profile for extruding polygon
[edge,z] = edgeprofile(4);
%  extrude polygon
p = tripolygon(poly,z,'edge',edge);
\end{verbatim}
The corresponding particle is displayed in the right panel of Fig.~\ref{fig:particleshapes}.  The command \verb|edgeprofile| returns an array of $(x,y)$ values that control the rounding-off of the particle.  With \verb|tripolygon| we triangulate the rounded 2D triangle and extrude the particle along the $z$ direction.  In the help pages of the toolbox we explain in more detail how to call \verb|tripolygon| and related functions in order to utilize the full potential of the \texttt{Mesh2d} toolbox.

\section{Dielectric environment}\label{sec:environment}

\subsection{Dielectric functions}

There exist three classes for dielectric functions.
\begin{description}

\item[\texttt{epsconst}] A constant dielectric function is initialized with \verb|epsconst(val)|, where \verb|val| is the dielectric constant.

\item[\texttt{epsdrude}] A Drude dielectric function for metals $\varepsilon(\omega)=\varepsilon_0-\omega_p^2/\omega(\omega+i\gamma)$ is initialized with \verb|epsdrude(name)|, where \verb|name| is the name of the metal.  We have implemented \verb|'Au'|, \verb|'Ag'|, and \verb|'Al'| for gold, silver, and aluminum.

\item[\texttt{epstable}] A tabulated dielectric function is initialized with \verb|epstable(finp)|, where \verb|finp| is the name of an input file.  This file must be in ASCII format where each line holds the values \verb|ene n k|, with \verb|ene| being the photon energy in eV, and \verb|n| and \verb|k| the real and imaginary part of the refractive index $\sqrt{\varepsilon}$, respectively.  In the toolbox we provide the files \verb|'gold.dat'| and \verb|'silver.dat'| for the gold and silver dielectric functions tabulated in Ref.~\cite{johnson:72}.

\end{description}
For a dielectric object \verb|objeps| of one of these classes, one can compute the dielectric function and the wavenumber inside the medium with
\begin{verbatim}
[eps,k] = epsobj(enei);
\end{verbatim}
where \verb|enei| is as usual the wavelength of light in vacuum.

\subsection{The {\rm\texttt{comparticle}} class}

The \texttt{comparticle} class defines how the particle boundaries are embedded in the dielectric environment.  As previously discussed, the outer surface normal $\bm n$ allows to distinguish between the boundary in- and out-side.  For more complex particles, such as a dumbbell-like particle, it is not alway possible to chose the particle boundaries such that only particle insides or outsides are connected.  In these cases the meaning of in- and out-side is just a matter of convention. 

To initialize a dielectric environment, one calls
\begin{verbatim}
p = comparticle({eps1,eps2,...},{p1,p2,...},inout,closed);
\end{verbatim}
The first and second argument are cell arrays of the dielectric functions and particle boundaries characterizing the problem.  In general, we recommend to set the first entry of the dielectric functions to that of the embedding medium, as several functions, such as for the calculation of the scattering or extinction cross sections, assume this on default.  The third argument \verb|inout=[i1,o1;i2,o2;...]| defines for each particle boundary \verb|p.| the dielectric functions \verb|eps{i.}| and \verb|eps{o.}| at the in- and out-side of the boundary.  For instance, for the particles depicted in Fig.~1 we get
\begin{verbatim}
%  single sphere of Fig. 1(a)
p = comparticle({eps1,eps2},{p},[2,1],1);
%  coupled spheres of Fig. 1(b)
p = comparticle({eps1,eps2,eps3},{p1,p2},[2,1;3,1],1,2);
%  coated particles of Fig. 1(c)
p = comparticle({eps1,eps2,eps3},{p1,p2},[2,1;3,2],1,2);
\end{verbatim}

The last argument (or arguments) of the \verb|comparticle| initialization define closed boundaries.  In general, for a closed boundary $\partial V_i$ the following sum rule applies \cite{fuchs:76}
\begin{equation}\label{eq:sumrule}
  \oint_{\partial V_i}\frac{\partial G(\bm s,\bm s')}{\partial n}\,da'=2\pi\,,\quad
  \sum_j\left(\frac{\partial G}{\partial n}\right)_{ij}=2\pi\,,
\end{equation}
which can be used to compute the diagonal elements of the surface derivative of the Green function.  Surprisingly, these diagonal elements play an extremely important role for obtaining accurate results even for coarse boundary discretizations.  For this reason it is important to indicate closed particle boundaries.  In the above examples we specify that boundaries \verb|p|, \verb|p1|, and \verb|p2| are closed.  For instance, when the boundary of a particle is composed of two objects \verb|p1| and \verb|p2| one indicates this with
\begin{verbatim}
p = comparticle({eps1,eps2},{p1,p2},[2,1;2,1],[1,2]);
\end{verbatim}
These calling sequences for \verb|comparticle| are sufficiently general to cope with even more complicated compositions of dielectric particles.

\subsection{The {\rm\texttt{compoint}} class}

The \texttt{MNPBEM} toolbox also provides a class \texttt{compoint} for points within a dielectric environment.  These objects are particularly useful for computing maps of electromagnetic fields or defining the properties of oscillating dipoles.  Consider a list of $n$ positions \verb|poslst|, i.e. a $n\times 3$ array, together with a \texttt{comparticle} object \verb|p|.  With
\begin{verbatim}
%  place the points into the dielectric environment
pt = compoint(p,poslst);
pt = compoint(p,poslst,'medium',1);
pt = compoint(p,poslst,'mindist',2);
\end{verbatim}
we can place these points inside the dielectric environment of \verb|p|.  By passing the additional property name \verb|'medium'| we can select the dielectric media within which points are kept, and with \verb|'mindist'| we can set a minimum distance between the points and the particle boundaries.  Below we give some examples for using \texttt{compoint} objects.

\subsection{The {\rm\texttt{compstruct}} class}

Internally, the \texttt{MNPBEM} toolbox stores the scalar and vector functions as \texttt{compstruct} objects, which behave very much like Matlab \texttt{struct} objects, but must always hold a \texttt{comparticle} or \texttt{compoint} object \verb|p| and the light wavelength \verb|enei|
\begin{verbatim}
%  set up a compstruct object
c = compstruct(p,enei);
%  add fields to the object
c.val = val;
\end{verbatim}
In many respects one can treat these objects as normal arrays.  This means, we can add or subtract \texttt{compstruct} objects and we can multiply them with a constant value.  In adding or subtracting them, fields that are missing in one of the objects are treated as zeros.  Upon multiplication all fields of the \texttt{compstruct} object are multiplied with the same value.  These features are particularly useful for \texttt{compstruct} object with electromagnetic potentials or fields that can be easily added or scaled.

\section{BEM solvers}\label{sec:solver}

The quasistatic BEM solver is initialized with
\begin{verbatim}
%  initialize BEM solver
bem = bemstat(p);
%  initialization for given wavelength enei
bem = bemstat(p,enei);
%  intialization passing arguments to the BEM solver
bem = bemstat(p,[],op);
\end{verbatim}
For the solution of the full Maxwell equations we simply have to use the solver \verb|bemret|.  When we pass the argument \verb|enei| to the BEM solver, the matrices of the BEM approach are computed for the given light wavelength.  Alternatively, we can compute the matrices later through
\begin{verbatim}
bem = bem(enei);
\end{verbatim}
For practically all BEM simulations this call will consume most of the computer time.

One can also pass options to the BEM solvers.  In general, the BEM approach implemented within the \texttt{MNPBEM} toolbox uses a \textit{collocation} scheme where the surface charges and currents $\sigma$ and $\bm h$ are assumed to be situated at the centroids of the boundary elements.  More accurate results could be achieved by performing a linear interpolation of $\sigma$ and $\bm h$ within the boundary elements.  However, so far we have refrained from such an interpolation because it would make the implementation of the BEM equations much more complicated.  In addition, in comparison with the field-based BEM approaches, which invoke second-order surface derivatives of the Green function, the collocation scheme for the potential-based BEM approach is expected to be of the same order of accuracy as a linear-interpolation scheme for a field-based BEM approach.  In some cases, e.g. when encountering the elongated surface elements of extruded particles, see Fig.~\ref{fig:particleshapes}, a pure collocation approach is problematic and it is better to assume that $\sigma$ and $\bm h$ are constant over the face elements.  We then have to integrate in $G_{ij}$ and its surface derivative over boundary elements $j$ which are sufficiently close to the element $i$.  In order to do so, one can define in the options
\begin{verbatim}
op = green.options('cutoff',cutoff);
\end{verbatim}
a \verb|cutoff| parameter that determines whether such a face integration is performed or the function value between the collocation points is taken.  As will be discussed below, we can pass \verb|op| also directly to \verb|compgreen| objects.

\subsection{Solving the BEM equations}

In the following we examine the working principle of the BEM solvers for the planewave excitation discussed in Sec.~\ref{sec:start}
\begin{verbatim}
%  plane wave excitation for given light polarizations
exc = planewavestat([1,0,0;0,1,0]);
%  planewave excitation
e = exc(p,enei);
\end{verbatim}
In the last call we receive a \texttt{compstruct} object \verb|e| which holds the field \verb|phip| for the surface derivative $\frac{\partial\phi_{\rm ext}}{\partial n}$ of the scalar potential.  Similarly, in the retarded case the returned object contains fields for the scalar and vector potentials at the boundaries, together with their surface derivatives.

For this external excitation we can now compute the surface charges through
\begin{verbatim}
%  initialize BEM solver
bem = bem(enei);
%  compute surface charge
sig = bem \ e;
\end{verbatim}
Alternatively, we can also put all commands into a single line
\begin{verbatim}
%  set up BEM solver
bem = bemstat(p);
%  initialize BEM solver and compute surface charge
sig = bem \ exc(p,enei);
\end{verbatim}
This calling sequence is simpler but it has the disadvantage that the BEM matrices, whose computation is rather time consuming, are not stored in \verb|bem| after the call.  If we are only interested in a single type of excitation this is not a problem, but if we want to compute surface charges for the same wavelength \verb|enei| but for a different excitation, e.g. dipole excitation, it is better to store the matrices through \verb|bem=bem(enei)|.

\begin{figure}
\centerline{\includegraphics[height=0.3\columnwidth]{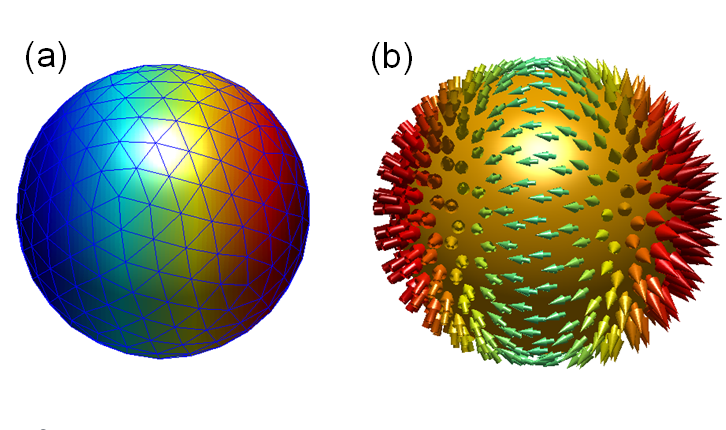}
            \includegraphics[height=0.3\columnwidth]{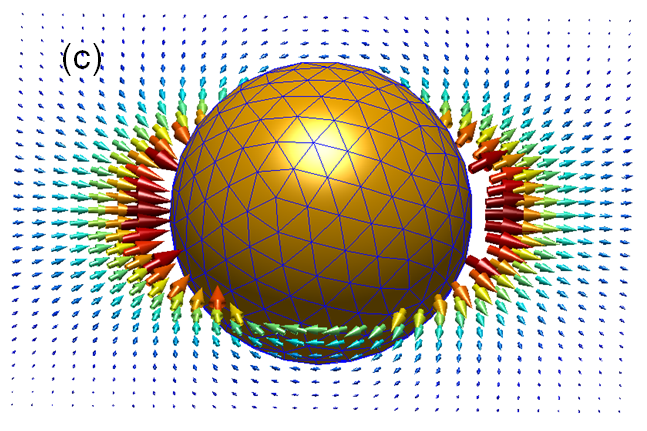}}
\caption{Results of BEM simulations: (a) surface charges $\sigma_i$, and real part of induced electric field at (b) particle boundary and (c) elsewhere.  In the simulations (consult the demo file \texttt{demofieldstat.m} for details) we consider a plane wave excitation of a gold nanoparticle with a diameter of 10 nm embedded in water.}\label{fig:bem}
\end{figure}            

We can now plot the surface charge \verb|sig.sig| and the electric field at the particle boundary through
\begin{verbatim}
%  plot real part of surface charge
plot(p,real(sig.sig),'EdgeColor','b');
%  electric field at outside of particle boundary
field = bem.field(sig,2);
%  plot real part of electric field
plot(p,'cone',real(field.e),'scale',0.6);
\end{verbatim}
The results are shown in panels (a) and (b) of Fig.~\ref{fig:bem}.

\subsection{Green functions}

Finally, we show how to proceed when we want to compute the electromagnetic fields or potentials elsewhere.  To this end, we have to set up a \texttt{compgreen} object, which remains usually hidden within the BEM solvers.  With \verb|pt| a \verb|compoint| object of the positions where the electromagnetic field should be computed and \verb|p| a \verb|comparticle| object for the particle boundaries, the Green function can be initialized with
\begin{verbatim}
%  set up Green function between points and particle
g = compgreen(pt,p);
%  same as above but with additonal cutoff parameter
g = compgreen(pt,p,green.option('cutoff',cutoff));
\end{verbatim}
With the help of the Green function we can compute according to Eq.~\eqref{eq:adhoc} the induced fields and potentials everywhere else.  Below we show the code needed in order to produce Fig.~\ref{fig:bem}(c)
\begin{verbatim}
%  regular mesh
[x,y] = meshgrid(linspace(-10,10,31));
pt = compoint(p,[x(:),y(:),0*x(:)],'mindist',1);
%  set up Green function between mesh points and particle
g = compgreen(pt,p);
%  compute electric field
field = g.field(sig);
%  plot particle and real part of electric field
plot(p,'EdgeColor','b');
coneplot(pt.pos,real(field.e),'scale',0.6);
\end{verbatim}

\section{BEM Simulations}\label{sec:simulation}

The \texttt{MNPBEM} toolbox provides two excitation-measurement schemes:  one for planewave excitation and the calculation of the scattering and extinction cross sections, which we have already discussed in Sec.~\ref{sec:start}, and one for the excitation of an oscillating dipole and the calculation of the enhancement of the radiative and total scattering rates. 

\subsection{Quasistatic versus full BEM simulations}

Let us look to the file \texttt{democputime.m}, available in the \texttt{Demo} subdirectory, which compares for the calculation of the scattering cross sections of a nanosphere the CPU times for different surface discretizations and for the different BEM solvers.  In addition to \texttt{bemstat} and \texttt{bemret} we also show results for the \texttt{bemstateig} solver, which solves the BEM equations using a restricted number of eigemodes of the matrix $\left(\frac{\partial G}{\partial n}\right)_{ij}$ \cite{fuchs:75,mayergoyz:05,truegler.prb:11}
\begin{verbatim}
CPU time elapsed for BEM simulations in seconds
#verts    #faces   bemstat  bemstateig    bemret
   144       284      1.07        0.14      9.31
   256       508      4.07        0.42     37.40
   400       796     14.01        1.26    129.57
   676      1348     62.24        2.97    651.25
\end{verbatim}
From the results it is apparent that the simulations based on the full Maxwell equations are about a factor of ten slower than those based on the quasistatic approximation.  An additional speedup can be achieved for the eigenmode expansion, in the above example we have used 20 eigenmodes.

The question which BEM solver is the most appropriate one cannot be answered in a unique way. 

\begin{description}

\item[Quasistatic solvers.]  The quasistatic solvers \texttt{bemstat} and \texttt{bemstateig} are ideal for testing and getting a feeling of how the results will approximately look like, at least for structures which are significantly smaller than the light wavelength.  It is a matter of taste what one calls ``significantly smaller'', but metallic spheres with diameters below say 50 nm and flat or elongated particles with dimensions below 100 nm will probably do. If you are dealing with even smaller structures, with dimensions of a few tens of nanometers, the quasistatic approximation will probably work perfectly in all cases.  However, we recommend to compare from time to time with the results of the full BEM solver \texttt{bemret}.

\item[Full BEM solver.]  BEM simulations based on the full Maxwell equations are much slower than those performed with the quasistatic BEM solvers, the main reason being the numerous matrix inversions.  For a given number $N$ of particle faces, the time needed for a matrix inversion is of the order $N^3$.  For this reason it is good to keep the number of faces and vertices as small as possible.  Nevertheless, in many cases it is indispensable to solve the full Maxwell equations.  Typical simulation times for the \texttt{bemret} solver are in the range between minutes and a few hours. 

\end{description}

\subsection{Dipole excitations}

\begin{figure}[t]
\centerline{\includegraphics[width=0.7\columnwidth]{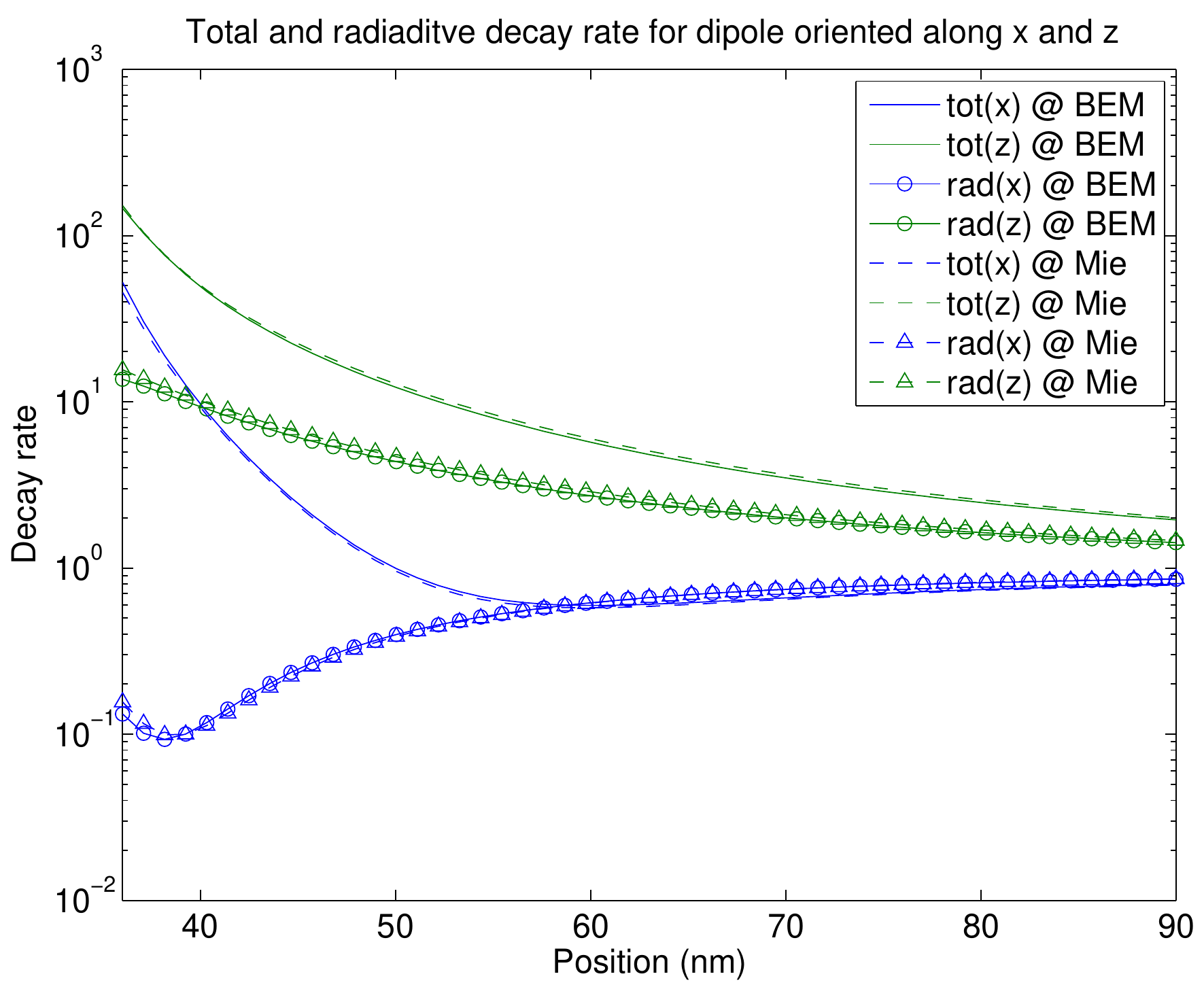}}
\caption{Enhancement of the total and radiative decay rates of a dipole located in the vicinity of a metallic nanosphere with a diameter of 60 nm.  The dipole is located on the $z$ axis at different distances from the sphere center.  We use a dielectric function representative for gold \cite{johnson:72} and a refractive index of $n_b=1.33$ for the embedding medium.  The results of the BEM simulation (\texttt{demospectrumret.m}) are almost indistinguishable from those of Mie theory.}\label{fig:dipole}
\end{figure}

We next discuss how to simulate the excitation of an oscillating electric dipole located at some distance from a nanoparticle.  This situation corresponds to the decay of an excited molecule or quantum dot, whose decay rates become enhanced through the modified photonic environment.  Similarly, the dipole excitation can be also used to compute the dyadic Green function of Maxwell's theory (see below).

The class \texttt{dipolestat} allows for an implementation of this problem.  Suppose that we have a \texttt{compoint} object \verb|pt| with the positions of the dipoles.  To implement a \texttt{dipolestat} object we call
\begin{verbatim}
%  default dipole orientations x, y, z
dip = dipolestat(pt);
%  dipole orientations x, z
dip = dipolestat(pt,[1,0,0;0,0,1]);
%  user-defined dipole vectors for each position
dip = dipolestat(pt,vec,'full');     
\end{verbatim}
Once we have set up the dipole excitation, we can use it in combination with the quasistatic BEM solvers (for the simulation of the full Maxwell equations we simply have to use \verb|dipoleret|, in order to compute both the scalar and vector potentials of the external dipole excitation, as well as \verb|bemret|).  The enhancement of the radiative and total decay rates, with respect to the free-space decay rate, is then computed with \cite{novotny:06,hohenester.ieee:08}
\begin{verbatim}
%  BEM simulation
sig = bem \ dip(p,enei );
%  enhancement of total and radiative decay rate
[tot,rad] = dip.decayrate(sig);
\end{verbatim}
Figure~\ref{fig:dipole} shows the results of \verb|demodipoleret.m| for an oscillating dipole located in the vicinity of a metallic nanosphere, as well as the comparison with Mie theory.  

Finally, through the relation
\begin{equation}
  \bm E(\bm r)=k^2\,\bm G(\bm r,\bm r';\omega)\cdot\bm d
\end{equation}
we are in the position to compute the dyadic Green function $\bm G(\bm r,\bm r';\omega)$ of Maxwell theory.  To this end we place an oscillating dipole at position $\bm r'$, solve the BEM equations for a given wavenumber $k=\omega/c$, and finally compute the electric field at the positions $\bm r$ according to the prescription given in Sec.~\ref{sec:solver}.

\subsection{Setting up a new excitation-measurement scheme}

In some situations one might like to set up a new excitation-measurement scheme, e.g., to simulate a nearfield excitation or electron energy loss spectroscopy (EELS).  Setting up a new excitation-measurement scheme can be done with a moderate amount of work.  In the following we sketch how this should be done.  Further information can be found in the help pages.

We recommend to use Matlab classes.  An object \verb|exc| of this class should return with \verb|exc(p,enei)| the potentials (and their surface derivatives) for the external perturbation at the particle boundary, which can be achieved by implementing the \verb|subsref| function.  To see how this can be done efficiently, we suggest to inspect the \verb|planewave| and \verb|dipole| classes of the toolbox.  As for the measurement implementation, it probably suffices to set up a \verb|compgreen| object and to compute with it the induced electromagnetic fields or potentials, which then can be further processed.

\section{Summary and outlook}

To summarize, we have presented a Matlab toolbox \texttt{MNPBEM} suited for the simulation of metallic nanoparticles (MNP) using a boundary element method (BEM) approach.  The toolbox relies on the concept of Matlab classes which can be easily combined, such that one can adapt the simulation programs flexibly for various applications.  All technicalities of the potential-based BEM approach remain hidden within the classes.  The toolbox provides detailed help pages and a collection of demo programs.  Several plot commands allow to access the full potential of the Matlab program and facilitate the analysis and interpretation of the simulation results.

As regarding simulation time and accuracy, we believe that the toolbox performs well and can compete with the other simulation toolkits used in the plasmonics community, although more detailed tests are needed for clarification.  There is plenty of room for improvements, such as multigrid methods or interpolation schemes beyond the presently used collocation scheme.  We are presently testing the implementation of mirror symmetry, which can speed up simulations by about one order of magnitude, as well as of particles placed on substrates or embedded in layer structures.  Both features are working fine, but we have not included them in this version of the toolbox because they are still somewhat experimental.  Future work will also address periodic structures and static electric fields, such as needed for electrochemistry.  Altogether, we hope that the \texttt{MNPBEM} toolbox will serve the plasmonics community as a useful and helpful simulation toolkit.

\section*{Acknowledgment}

We thank the \texttt{nanooptics} group in Graz for most helpful discussions.  This work has been supported in part by the Austrian science fund FWF under project P21235--N20.





\bibliographystyle{elsarticle-num}

\end{document}